\documentclass{sig-alternate}

\usepackage{epsfig}
\newtheorem{theorem}{Theorem}[section]
\newtheorem{corollary}{Corollary}[section]
\newtheorem{lemma}{Lemma}[section]
\newtheorem{proposition}{Proposition}[section]
\newtheorem{definition}{Definition}[section]
\newtheorem{non-theorem}{Non-theorem}[section]

\newcommand{\thm}{\begin{theorem}}

\newcommand{\lem}{\begin{lemma}}

\newcommand{\pro}{\begin{proposition}}

\newcommand{\dfn}{\begin{definition} \rm}

\newcommand{\rem}{\begin{remark}}

\newcommand{\xam}{\begin{example}}

\newcommand{\cor}{\begin{corollary}}
\newcommand{\prf}{\begin{proof}}

\newcommand{\ethm}{\end{theorem}}

\newcommand{\elem}{\end{lemma}}

\newcommand{\epro}{\end{proposition}}

\newcommand{\edfn}{\bbox\end{definition}}

\newcommand{\erem}{\bbox\end{remark}}

\newcommand{\exam}{\bbox\end{example}}

\newcommand{\ecor}{\end{corollary}}

\newcommand{\eprf}{\end{proof}}

\newcommand{\beqn}{\begin{equation}}

\newcommand{\eeqn}{\end{equation}}

\newcommand{\bbox}{\vrule height7pt width4pt depth1pt}

\newcommand{\commentout}[1]{}

\newenvironment{RETHM}[2]{\trivlist \item[\hskip 10pt\hskip\labelsep{\sc #1\hskip 5pt\relax\ref{#2}.}]\it}{\endtrivlist}
\newcommand{\rethm}[1]{\begin{RETHM}{Theorem}{#1}}
\newcommand{\repro}[1]{\begin{RETHM}{Proposition}{#1}}
\newcommand{\relem}[1]{\begin{RETHM}{Lemma}{#1}}
\newcommand{\recor}[1]{\begin{RETHM}{Corollary}{#1}}

\newcommand{\erethm}{\end{RETHM}}
\newcommand{\erepro}{\end{RETHM}}
\newcommand{\erelem}{\end{RETHM}}
\newcommand{\erecor}{\end{RETHM}}
\newcommand{\BR}{\mathit{BR}}

\begin{document}
\conferenceinfo{EC'07,} {June 13--16, 2007, San Diego, California, USA.}
\CopyrightYear{2007}
\crdata{978-1-59593-653-0/07/0006}

\title{Optimizing Scrip Systems: Efficiency, Crashes, Hoarders, and
Altruists}

\numberofauthors{3}

\author{
\alignauthor{Ian A. Kash}\\
\affaddr{Computer Science Dept.}\\
\affaddr{Cornell University}\\
\email{kash@cs.cornell.edu}
\alignauthor{Eric J. Friedman}\\
\affaddr{School of Operations Research and Industrial Engineering}\\
\affaddr{Cornell University}\\
\email{ejf27@cornell.edu}
\alignauthor{Joseph Y. Halpern}\\
\affaddr{Computer Science Dept.}\\
\affaddr{Cornell University}\\
\email{halpern@cs.cornell.edu}
}

\date{ }
\maketitle

\begin{abstract}
We discuss the design of efficient scrip systems and
develop tools for empirically analyzing them.
For those interested in the empirical study of scrip systems,

we demonstrate how characteristics of agents in a system can be
inferred from the equilibrium distribution of money. From the
perspective of a system designer, we examine the effect of the money
supply on social welfare and show that social welfare is maximized
by increasing the money supply up to the point that the system
experiences a ``monetary crash,'' where money is sufficiently
devalued that no agent is willing to perform a service. We also
examine the implications of the presence of altruists and hoarders
on the performance of the system.  While a small number of altruists
may improve social welfare, too many can also cause the system to
experience a monetary crash, which may be bad for social welfare.
Hoarders generally decrease social welfare but, surprisingly, they
also promote system stability by helping prevent monetary crashes.
In addition, we provide new technical tools for analyzing and
computing equilibria  by showing that our model exhibits strategic
complementarities, which implies that there exist equilibria in pure
strategies that can be computed efficiently.
\end{abstract}




\section{Introduction}\label{sec:intro}

Historically, non-governmental organizations have issued their own
currencies for a wide variety of purposes.  These currencies, known as
\emph{scrip}, have been used in company towns where government issued

currency was scarce \cite{mining},  in Washington DC to reduce the
robbery rate of bus drivers \cite{busdrivers}, and in Ithaca (New York) to
promote fairer pay and improve the local economy \cite{hours}.
Scrip systems are also becoming more prevalent in online systems.
To give just some examples,
the currencies of online virtual worlds such as Everquest and
Second Life are a form of scrip;
market-based solutions
using scrip systems
have been suggested

for applications such as system-resource allocation \cite{agora},
managing replication and query

optimization in a distributed database \cite{mariposa}, and allocating
experimental time on a wireless sensor network test bed \cite{mirage};
a number of sophisticated scrip
systems have been proposed \cite{gupta03,fileteller02,karma03} to allow
agents to pool resources while
avoiding what is known as \emph{free riding}, where agents
take advantage of the resources the system provides while providing none
of their own (as Adar and Huberman \citeyear{adar00} have shown, this
behavior certainly takes place in systems such as Gnutella);
and
Yootles \cite{yootles} uses a scrip
system as a way of helping groups make decisions
using economic mechanisms
without involving real
money.

Creating a scrip system creates a new market where goods and services
can be exchanged that may have been impractical or undesirable to
implement with standard currency.
However, the potential benefits of a scrip system will not necessarily
be realized simply by creating the framework to support one.  The
story of the Capitol Hill Baby Sitting Co-op \cite{babysitting},
popularized by Krugman \cite{Krugman}, provides a cautionary tale.
The Capitol Hill Baby Sitting Co-op was a group of parents working on
Capitol Hill who agreed to cooperate to provide babysitting services
to each other.  In order to enforce fairness, they issued a supply of
scrip with each coupon worth a half hour of babysitting.  At one
point, the co-op had a recession.  Many people wanted to save up
coupons for when they wanted to spend an evening out.  As a result,
they went out less and looked for more opportunities to babysit.
Since a couple could earn coupons only when another couple went out,
no one could accumulate more, and the problem only got worse.

After a
number of failed attempts to solve the problem, such as mandating a
certain frequency of
going out, the co-op started issuing more coupons.  The results were striking.
Since couples had a sufficient reserve of coupons, they were more
comfortable spending them.  This in turn made it much easier to earn
coupons when a couple's supply got low.
Unfortunately, the amount of scrip grew to the point
that most of the couples felt ``rich.''  They had
enough scrip for the foreseeable future, so naturally they didn't want
to devote their evening to babysitting.  As a result,
couples who wanted to go out were unable to find another couple willing
to babysit.

This anecdote shows that the amount of scrip in circulation can
have a significant impact on the effectiveness of a scrip
system.
If there is too little money in the system, few agents will be
able to afford service.
At the other extreme, if
there is too much money in the system, people
feel rich and stop looking for work. Both of these extremes lead to
inefficient outcomes.  This suggests that there is an optimal amount
of money, and that
nontrivial deviations from the optimum towards either extreme
can lead to significant degradation in the performance of the system.

Motivated by these observations,
we study the behavior of scrip systems with a
heterogeneous population of agents.  We prove a number of theoretical

results, and use them to gain
insights into the design and analysis of practical scrip systems.

The
motivation for our
interest in heterogeneous populations of agents should be clear.
In the babysitting coop example, we would not expect all couples to feel
equally strongly about going out nor to feel the ``pain'' of babysitting
equally.
In earlier work \cite{scrip06}, we showed that with a homogeneous
population of agents, we could assume that all agents were following a
\emph{threshold strategy}: an agent who has more than a certain threshold of
money will not volunteer to work; below the threshold, he will
volunteer.  Perhaps not surprisingly,
we show that
even with a heterogeneous
population, each different type of agent can still be characterized by a
threshold (although different types of agents will have different
thresholds).

We also characterize
the
distribution of money in the system in equilibrium,
as a function of the fraction of agents of each type.  Using this
characterization,
we show that we can infer the threshold strategies that
different types of agents are using simply from the distribution of
money.  This shows that, by simply
observing a scrip system in operation, we can learn a great deal about
the agents in the system.  Not only is such information of interest to
social scientists and marketers, it is also important to a system
designer trying to optimize the performance of the system.
This is because
agents that have no money will be unable to pay for service;
agents that are at their threshold are unwilling to serve
others.
We show that, typically,
it is the number of agents with no money that has the more
significant impact on the overall efficiency of the system.
Thus, the way to optimize the performance of the system is to try to
minimize the number
of agents with no money.

As we show, we can decrease the number of agents with no money by
increasing the money supply.
However, this only works up to a point.
Once
a certain
amount of money is reached, the system experiences a monetary crash:
there is so much money that, in equilibrium, everyone will feel rich and
no agents are willing to work.
The point where the system crashes gives us a sharp threshold.
We show that,
to get optimal performance, we want the total amount of money in the
system to be as close as possible to the threshold, but not to go over
it.  If the amount of money in the system is over threshold,
we get the worst possible equilibrium, where all
agents have utility 0.

The analysis above assumes that all users have somewhat similar
motivation: in particular, they do not get pleasure simply from
performing a service, and are interested in money only to the extent
that they can use it to get services performed.
But in real systems, not all agents
have this motivation.
Some of the more common ``nonstandard'' agents are \emph{altruists} and
\emph{hoarders}.
Altruists are willing to satisfy all requests,
and do not require money in return; hoarders never
make requests, and just hoard the money they make by satisfying
requests.
Studies of the Gnutella peer-to-peer file-sharing network have shown
that one percent of agents satisfy fifty percent of the requests
\cite{adar00}.  These agents are doing significantly more work for
others than they will ever have done for them,
so can be viewed as altruists.
Anecdotal evidence also suggests that the introduction of any sort of
currency seems to inspire hoarding behavior on the part of some agents,
regardless of the benefit of possessing money.

 Altruists and hoarders have opposite effects on a system: having
altruists is essentially equivalent to adding money; having hoarders
is essentially equivalent to removing it.
With enough altruists in the system, the system has a monetary crash,
in the sense
that standard agents stop being willing
to provide service,
just as when there is too much money in the system.
We show that, until we get to the point where the
system crashes,
the utility of
standard agents is improved by the presence of altruists.
However, they can be worse off in a system
that experiences a monetary crash due to the presence of many
altruists than they would be if there were no altruists at all.
Similarly,
we show that, as the fraction of
hoarders
increases, standard agents begin to suffer because there is
effectively less money in circulation.
On the other hand, hoarders can improve the
stability of a system.  Since hoarders are willing to accept an
infinite amount of money, they can
prevent the monetary crash that would
otherwise occur as money was added.
In any case, our results show that the presence of both altruists and
hoarders can be
mitigated by appropriately controlling the money supply.

In order to examine these issues, we
use a model of a scrip system
that we developed in previous work
\cite{scrip06}.  While the model was developed with the workings of a
peer-to-peer network in mind
and assumed that all agents were identical,
the model applies to a wide variety of scrip systems, and makes perfect
sense even with a heterogeneous population of agents.
We showed that, under
reasonable assumptions, a system with only one type of agent has
a cooperative equilibrium using threshold strategies.
Here we extend the theoretical results to the case of multiple types
of agents. We also introduce a new argument for the existence of
equilibria that relies on the monotonicity of the best-reply
function.
We show that if some agents change their strategy to one with a higher
threshold, no other agent can do better by lowering his threshold.

This makes our game one with what Milgrom and Roberts
\citeyear{MiR90} call \emph{strategic complementarities},

who (using the results of Tarski \cite{tarski} and Topkis
\cite{topkis}) showed that there are pure strategy equilibria in

such games, since the process of starting with a strategy profile
where everyone always volunteers (i.e., the threshold is $\infty$)
and then iteratively computing the best-reply profile to it

converges to a Nash equilibrium in pure strategies. (Our earlier
results guaranteed only an equilibrium in mixed strategies.) This
procedure also provides an efficient algorithm for explicitly
computing equilibria.

The rest of the paper is organized as follows.  In Section
\ref{sec:model}, we review our earlier model.
In Section \ref{sec:theory}, we
prove basic results about the
existence and form of equilibria.
Sections \ref{sec:distribution},
\ref{sec:optimize}, and \ref{sec:altruists} examine the practical
implications of our theoretical results.  Section \ref{sec:distribution}
examines the distribution of money in the system.
We give an explicit formula for the distribution of money in
the system based, and
show how it can be used to infer
the number of types of agents
present and the strategy each type is using.
In Section \ref{sec:optimize},
we examine how a system designer can optimize the performance of the
system by adjusting the money supply appropriately.
Section \ref{sec:altruists} examines how altruists and hoarders
affect
the system.  We conclude in Section~\ref{sec:conclusion}.

\section{Our Model} \label{sec:model}

We begin by reviewing our
earlier
model of a scrip system with $n$ agents.
In the system, one agent can request a service which another
agent can volunteer to fulfill.  When a service is performed by agent
$i$ for agent $j$, agent $i$ derives some utility from having that
service performed, while agent $j$ loses some utility for performing
it.  The amount of utility gained by having a service performed and the
amount lost be performing it may depend on the agent.
We assume that agents have a \emph{type} $t$ drawn from some finite set
$T$ of types.

We can describe the entire population of agents
using the triple
$(T,\vec{f},n)$, where $f_t$ is the fraction with type $t$ and $n$ is the
total number of agents.
For most of the paper, we consider only what we call \emph{standard
agents}.  These are agents who derive no pleasure from performing a
service, and for whom money has no intrinsic value.  We can characterize
the type of an agent by a tuple
$t = (\alpha_t, \beta_t, \gamma_t, \delta_t, \rho_t)$, where
\begin{itemize}
\item $\alpha_t$ reflects the cost of satisfying the request;
\item $\beta_t$ is the probability that the agent can satisfy the request
(an agent may not be able to satisfy all requests; for example, in a

babysitting co-op, an agent may not be able to babysit every night);
\item $\gamma_t$ measures the utility an agent gains for having a request
satisfied;
\item $\delta_t$ is the rate at which the agents discounts utility (so a
unit of utility in $k$ steps is worth only $\delta^k$ as much as a unit
of utility now)---intuitively, $\delta_t$ is a measure of an agent's
patience (the larger $\delta_t$ the more patient an agent is, since a
unit of utility tomorrow is worth almost as much as a unit today); and
\item $\rho_t$ represents the (relative) request rate (since not all
agents make requests at the same rate)
---intuitively, $\rho_t$ characterizes an agent's ``neediness''.
\end{itemize}

We model the system as running for an infinite number of rounds.
In each round, an agent is picked with probability proportional to
$\rho$ to request service.  Receiving service costs some amount of scrip
that we normalize to \$1.

If the chosen agent does not have enough
scrip, nothing will happen in this round.  Otherwise,
each agent of type $t$ is able to satisfy this request with
probability $\beta_t$, independent of previous behavior.
If at least one agent is able and willing to satisfy the request,
and the requester has type $t$, then the requester gets a benefit of
$\gamma_t$ utils (the job is done) and
one
of the volunteers is chosen at random to fulfill the request.
If the chosen volunteer has type $t'$, then
that
agent pays a cost of $\alpha_{t'}$ utils, and receives a dollar as payment.
The utility of all other agents is unchanged in that round.
The total utility of an agent is the discounted sum of round utilities.
To model the fact that requests will happen more frequently the more
agents there are,
we assume that the time between rounds is
$1/n$.

This captures the intuition that things are really happening in
parallel and that adding more agents should not change an agent's
request rate.

One significant assumption we have made here is 
that prices are fixed.  While there are many systems with standard
``posted'' prices (the babysitting co-op is but one of many examples),
it certainly does not always hold in practice.  However, given the
potential costs of negotiating prices in a large system,

it does not seem so unreasonable to assume fixed prices.  Fixed prices
have the added advantage of making 
the analysis of agent 
strategies simpler, because the an agent knows the future cost of

requests rather than it being set as part of the equilibrium and

potentially varying over time.  We discuss this issue further
at the end of
Section \ref{sec:optimize}.

For more discussion of this model

and its assumptions, see \cite{scrip06}.

Our framework allows agents to differ in a number of parameters.
Differences in the parameters $\alpha$, $\gamma$, and $\delta$ seem easier
to deal with than differences in the other parameters
because they do not affect the action of the system except through
determining the optimal strategy.
We refer
to a population of types that differs only in these parameters as
one that exhibits only
\emph{payoff heterogeneity}.
Most of our results consider only payoff heterogeneity.
We do not believe that variation $\beta$ or $\rho$
fundamentally changes our results;
however, our existing techniques are
insufficient to analyze this case.
There is a long history of work in the economics literature on
the distribution of wealth dating back to the late 19th century
\cite{pareto}, 

although this work does not consider the distribution of money in the
particular setting of interest to us.

Hens et al.~\cite{hens} consider a related model.
There are a number of differences between the models.  First, in the Hens
et al.~model, there is essentially only one type of agent, but an
agent's utility for providing service (our $\gamma_t$) may change over
time.  Thus, at any time, there will be agents who differ in their
utility.  At each round, we assume that one agent is chosen (by nature)
to make a request for service, while other agents decide whether or not
to provide it.  In the Hens et al.~model,  at each round, each agent
decides whether to provide service, request service, or opt out, as a
function of his utilities and the amount of money he has.  They assume
that there is no cost for providing service and everyone is able to
provide service (i.e., in our language, $\alpha_t = 0$ and $\beta_t =
1$).  Under this assumption,
a system with optimal performance is one where half the agents request
service and the other half are willing to provide it.

Despite these differences, Hens et al.~also show that agents will use a
threshold strategy.   However, although they have inefficient
equilibria, because there is no cost for providing
service, their model does not exhibit the monetary crashes that our
model can
exhibit.

\section{Theoretical Results} \label{sec:theory}

In this section, we derive several basic results that provide insight
into the behavior of scrip systems with a heterogeneous population of
agents.
We first show that we can focus on a particularly simple class of
strategies:
\emph{threshold strategies}.

The strategy $S_k$ is the one in which the agent volunteers if and
only if his current amount of money is less than some fixed
threshold $k$. The intuition behind using a threshold strategy is
easy to explain: A rational agent with too little money will be
concerned that he will run out and then want to make a request; on
the other hand, a rational agent with plenty of money would not want
to work, because by the time he has managed to spend all his money,
the util will have less value than the present cost of working.  By
choosing an appropriate threshold, a rational agent can deal with
both concerns.

In \cite{scrip06}, we showed that if there is only one type of agent, it
suffices to consider only threshold strategies:
we show that (under certain mild assumptions) there exists a nontrivial
equilibrium where all agents use the same threshold strategy.
Here, we extend this result
to the case of payoff-heterogeneous agents.
To prove this result, we extend the characterization of the distribution
of money in a system where each agent uses the threshold strategy

provided in Theorem 3.1 of \cite{scrip06}.
To understand the characterization, note that
as agents spend and earn money, the distribution of money in the
system will change over time.  However, some distributions will be
far more likely than others.
For example, consider a system with
only two dollars.  With $n$ agents, there are $O(n^2)$ ways to assign
the dollars to different agents and $O(n)$ ways to assign them to the
same agent.
If each way of assigning the two dollars to agents is equally likely, we
are far more likely to see a distribution of money where two agents have
one dollar each than one
where a single
agent has two dollars.
It is well known \cite{jaynes} that
the distribution which has the most ways of being realized
is the one that maximizes entropy.
(Recall that the entropy of a probability distribution on a finite space
$S$ is  $-\sum_{s \in S} \mu(s) \log (\mu(s))$.)

Note that many distributions have no way of being realized.  For
example if the average amount of money available per agent is \$2
(so that if there are $n$ agents, there is $\$2n$ in the system),
then the distribution
where every agent has 3 dollars is impossible.  Similarly, if every
agent is playing $S_3$, then

a distribution that has some fraction of agents with \$4 is
impossible. 
Consider a scrip system where a fraction $\pi_k$ use strategy $S_k$.
(We are mainly interested in cases where $\pi_k = 0$ for all but finitely
many $k$'s, but our results apply even if countably many different
threshold strategies are used.)
Let $M^k_i$ be
the fraction of agents that play $S_k$ and have $i$ dollars.
Then the system must satisfy the following 

two
constraints:
\begin{eqnarray}
\sum_{k=0}^\infty \sum_{i = 0}^k i M^k_i &= &m \label{eqn:constraint1}\\
\sum_{i = 0}^k M^k_i &= &\pi_k \mbox{ for each $k$.} \label{eqn:constraint2}
\end{eqnarray}

These constraints capture the requirements that 
(\ref{eqn:constraint1}) the average amount of money is $m$ and 
(\ref{eqn:constraint2}) a fraction $\pi_k$ of the agents play $S_k$.
As the following theorem shows, in equilibrium, a large system is
unlikely to have a distribution far from the one that maximizes
entropy subject to
these constraints.

\thm \label{thm:maxent}
Given a payoff-heterogeneous system with $n$ agents where a fraction
$\pi_k$ of agents plays strategy $S_k$ and
the average amount of money is $m$,
let $M_{\vec{\pi},n,m}(t)$ be the 

the random variable that gives the
distribution of money
in the system at time $t$, and let
$M^*_{\vec{\pi},m}$ be the distribution that maximizes entropy subject to
constraints
(\ref{eqn:constraint1}) and (\ref{eqn:constraint2}).  Then
for all $\epsilon$,
there exists $n_\epsilon$ such that, for all $n > n_\epsilon$, there
exists a time $t^*$ such that for all $t > t^*$,
$$\Pr(||M_{\vec{\pi},n,m}(t) - M^*_{\vec{\pi},m}||_2 > \epsilon) <
\epsilon.
\footnote{Since $M_{\vec{\pi},n,m}(t)$ and $M^*_{\vec{\pi},m}$
are finite distributions, they can be viewed as vectors; $|| \cdot ||_2$
is then the standard $L_2$ distance between vectors.}
$$
\ethm

\prf (Sketch)
This theorem is proved for the homogenous case as Theorem 3.1 of
\cite{scrip06}.  Most of the proof applies without change to a
payoff-heterogeneous population, but one key piece differs.
This piece
involves showing that each possible assignment of money to agents is
equally likely;
this makes maximum entropy an accurate description of the likelihood of
getting a particular distribution of money.  We 

now
prove this by
considering the Markov chain whose
states are the possible assignments of dollars to agents
and whose transitions correspond to the possible
outcomes of a round, and showing that it has a uniform limit
distribution.
\lem \label{lem:stationary}
Given a scrip system with payoff-heterogeneous population $(T,\vec{f},n)$.
where all agents play some (possibly different) threshold strategy,
the limit distribution of the Markov chain that describes the
evolution of the system is the uniform distribution.
\elem

\prf
A sufficient condition for the limit distribution to be uniform is that
for every pair of states $s$ and $s'$, $P_{ss'} = P_{s's}$ (where

$P_{ss'}$ is the probability of transitioning from $s$ to $s'$).

If
$P_{ss'} > 0$, there must be some pair of agents $u$ and $v$ such that

$u$ has one more dollar in state $s$ than he does in $s'$, while $v$ has
one more dollar in $s'$ than in $s$.
Every other agent must have the same amount of money
in both $s$ and $s'$.
The key

observation is that, in both $s$ and $s'$, every agent other than $u$
and $v$ will make the same decision about whether to volunteer in each
state.
Additionally, in state $s$, $v$ is willing to volunteer if $u$ is
selected to make a request while in state $s'$, $u$ is willing to
volunteer if $v$ is selected to make a request.  An explicit
calculation of the probabilities shows that this means that $P_{ss'} =
P_{s's}$.

(Note that the last step in the lemma is where payoff heterogeneity
is important.

If $u$ is of type $t$, $v$ is of type $t'$, and either $\beta_t \neq
\beta_{t'}$ or $\rho_t \neq \rho_{t'}$, then it 
will, in general, not be the the case that $P_{ss'} = P_{s's}$.)

\eprf

\eprf 

Theorem \ref{thm:maxent} tells us that we can generally expect the
distribution of money to be close to the distribution that maximizes
entropy.  We can in fact give an exact characterization of this
distribution.

\cor \label{cor:maxent}

$(M^*_{\vec{\pi},m})^k_i = \pi_k \lambda^i / \sum_{j = 0}^k \lambda^j$ where
$\lambda$ is chosen to ensure that (\ref{eqn:constraint1}) is satisfied.
\ecor

\prf
The distribution we are looking for is the one that maximizes entropy
subject to (\ref{eqn:constraint1}) and (\ref{eqn:constraint2}).  This
means 

that
we want to
maximize $$\sum_{k} \sum_{i = 0}^k - M^k_i \log
M^k_i$$
subject to (\ref{eqn:constraint1}) and
(\ref{eqn:constraint2}).
Standard techniques,
using Lagrange multipliers \cite{jaynes}, show that $M^k_i$
must be of the given form.
\eprf

We now show that agents
have best responses among threshold strategies.

\thm \label{thm:threshold}
For all $m$, there exist $\delta^*$ and $n^*$ such that if
$(T,\vec{f},n)$ is a payoff-heterogeneous population with $n > n^*$ and
$\delta_t > \delta^*$ for all types $t \in T$,
then if each type $t$ plays some threshold strategy $S_{k_t}$
then every

agent of type $t$ has an $\epsilon$-best reply
\footnote{In \cite{scrip06}, we simply described this as a best
reply rather than an $\epsilon$-best reply, and noted that it might not
be a best reply if the distribution
is far from the maximum entropy distribution (which we know is very
unlikely).  Considering $\epsilon$-best replies and $\epsilon$-Nash
equilibria formalizes this intuition.}
 of the form $S_{k_t'}$.
Furthermore, either $k_t'$ is unique or
there are two best replies, which have the form
$k_t'$ and $k_t'+1$ for some $k_t'$.
\ethm

\prf This was proved for the homogeneous case as Theorem 4.1 of
\cite{scrip06}. The proof for the heterogeneous case is literally
identical, except that we use

Theorem \ref{thm:maxent} in place of the
analogous result for the homogeneous case.
\eprf

Theorem~\ref{thm:threshold} and Corollary~\ref{cor:maxent} assume that
all agents are playing threshold strategies; we have not yet shown that
there is a nontrivial
equilibrium where agents do so (all agents playing $S_0$ is a trivial
equilibrium). Our previous approach to proving the
existence of equilibria was to make the space of
threshold
strategies continuous.

For example, we considered strategies such as $S_{5.6}$, where the agent
plays $S_6$ with probability 0.6 and $S_5$ with probability 0.4.

We could then use standard fixed point techniques.
We believe that these arguments can be extended to the

payoff-heterogeneous case,
but we can in fact show more.

Our experiments showed that, in practice, we could always find
equilibria in pure strategies.  As we now show, this is not just
an artifact of the agent types we examined.

Given a payoff-heterogeneous population, let $\vec{k}$ denote the
strategy profile where type $t$ plays the threshold strategy $S_{k_t}$.
Let $\BR_{(T,\vec{f},n),m}^t(\vec{k})$ be the best reply for an agent of
type $t$ given that the population is
$(T,\vec{f},n)$, the average amount of money is $m$, and the strategy
profile is $\vec{k}$.
By Theorem \ref{thm:threshold}, for sufficiently large
$n$, this threshold is independent of $n$ and is either
unique or consists of two adjacent strategies; in the latter case, we
take $\BR_{(T,\vec{f},n),m}^t(\vec{k})$ to be the smaller of the two values.
We use
$\BR_{(T,\vec{f}),m}^t(\vec{k})$ to denote this $n$-independent unique best
response.

\lem \label{lem:monotone}

For all $m$ there exist $\delta^*$ and $n^*$ such that, if
$(T,\vec{f},n)$ is a payoff-heterogeneous population with $n > n^*$ and

$\delta_t > \delta^*$ for all $t$,
then the function
$\BR_{(T,\vec{f}),m}^t(\vec{k})$ is non-decreasing.
\elem

\prf (Sketch)
The population $(T,\vec{f},n)$ and $\vec{k}$ induce a distribution
$\vec{\pi}$ over strategies.
It is not hard to show that if $\vec{k}' \ge \vec{k}$ (i.e., $k'_t \ge
k_t$ for all types $t \in T$), then
$(M^*_{\vec{\pi},m})^{k'_t}_0 \ge (M^*_{\vec{\pi},m})^{k_t}_0$ and
$(M^*_{\vec{\pi},m})^{k'_t}_{k'_t} \le
(M^*_{\vec{\pi},m})^{k_t}_{k_t}$ for all types $t$.
This means that,
with $\vec{k}'$,

more agents will have zero dollars and be unable to afford a download,
and fewer agents will be at their threshold amount of money.
As a consequence, with $\vec{k}'$ there will be
fewer
opportunities to earn money and more agents wishing to volunteer for
those opportunities that do exist.  This
means that agents will earn money less often while wanting to spend
money at least as often (more volunteers means there is more likely to
be someone able to satisfy a request).
Therefore, with $\vec{k}'$, agents will run out of
money sooner.  Thus the value of earning an extra dollar increases and
so the best reply can only increase.
\eprf

\thm \label{thm:equilib}

For all $m$ there exist $\delta^*$ and $n^*$ such that, if
$(T,\vec{f},n)$ is a payoff-heterogeneous population with $n > n^*$ and

$\delta_t > \delta^*$ for all $t$,
then there exists a nontrivial $\epsilon$-Nash
equilibrium where all agents of type $t$ play $S_{k_t}$ for some
integer $k_t$.
\ethm

\prf (Sketch)
Let $\BR_{(T,\vec{f}),m}(\vec{k})$ be the strategy profile $\vec{k}'$
such that $k'_t = \BR^t_{(T,\vec{f}),m}(\vec{k})$.
By Lemma \ref{lem:monotone},
$\BR_{(T,\vec{f}),m}$ is non-decreasing,

so Tarski's fixedpoint
theorem \cite{tarski} guarantees the existence of a
greatest and least fixed point; these fixed points are equilibria.
The least fixed point is the trivial equilibrium.

We can compute the greatest fixed point by starting with the
strategy profile $(\infty, \ldots, \infty)$ (where each agent uses
the strategy $S_\infty$ of always volunteering) and considering
\emph{best-reply dynamics}, that is, iteratively computing the best-reply
strategy profile.  This process converges to the greatest fixed
point, which is an equilibrium (and is bound to be an equilibrium
in pure strategies, since the best reply is always a pure
strategy).
Furthermore, it is not hard to show that there exists some $\delta^*$ such
that if $\delta_t \ge
\delta^*$ for all types $t$, then

there exists a strategy profile $\vec{k}$ such 
that $\BR_{(T,\vec{f}),m}(\vec{k}) \ge \vec{k}$.
Monotonicity guarantees that
if $\vec{k}^*$ is the greatest fixed point of $\BR^{t}_{(T,\vec{f}),m}$,
then $\vec{k}^* \ge \vec{k}$.  Thus, the greatest fixed point gives a
nontrivial equilibrium.
\eprf

The proof of Theorem \ref{thm:equilib} also provides an algorithm for
finding equilibria that seems efficient in practice.  Start with the
strategy profile $(\infty, \ldots, \infty)$ and iterate the best-reply
dynamics until an equilibrium is reached.

\begin{figure}[htb]

\centering \epsfig{file=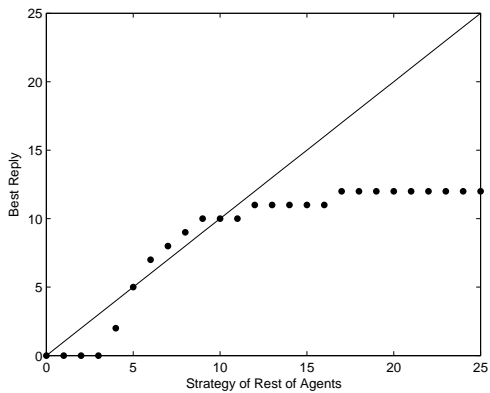, height=2.2in}
\caption{A hypothetical best-reply function with one type of agent.}
\label{fig:br}
\end{figure}

There is a subtlety in our results.
In general, there may be many equilibria.  From the perspective of

social welfare, some will be better than others.  As we show in
Section \ref{sec:optimize}, strategies that use smaller (but nonzero)
thresholds increase social welfare.
Consider the best-reply function with a single type of agent shown in
shown in Figure \ref{fig:br}.  An equilibrium must have $\BR(k) = k$,
so will be a point on the line $y = x$.  This example has three
equilibria: $S_0$, $S_5$, and $S_{10}$; $S_5$ is the equilibrium that
maximizes social welfare.
However, we cannot use best-reply dynamics to get to $S_5$, unless we
start there.
Applying best-reply dynamics to
a starting point above 10 will lead to convergence at $S_{10}$;
this is also true if we start at a point between 5 and 10.
On the other hand, starting at a point below 5 will lead to convergence
at $S_0$, the trivial equilibrium.
Thus, although $S_5$ is a more efficient equilibrium than $S_{10}$, it
is unstable.
The only equilibrium that we can guarantee is stable is the maximum one
(i.e., the greatest fixed point); thus, we focus on this equilibrium
in the rest of the paper.

\section{Identifying User Strategies} \label{sec:distribution}

In Section~\ref{sec:theory}, we used maximum entropy to get an
explicit formula for the distribution of money given
the fraction $\pi_k$ of agents using each strategy $S_k$, $k \ge 0$:
$M^k_i = \pi_k \lambda^i / \sum_{j = 0}^k \lambda^j$.
In this section, we want to go in the opposite direction:
given the distribution of money, we want to infer the fraction $f_k$
of agents using each strategy $S_k$, for each $k$.
For those interested in studying the agents of a scrip system,
knowing the fraction of agents using each strategy can provide a window
into the preferences of those agents.
For system designers, this knowledge is useful because, as we
show in Section~\ref{sec:optimize}, how much money the system can
handle without crashing depends on the fraction of agents of each type.

In equilibrium, the distribution of money has the form
described in
Corollary \ref{cor:maxent}.

Note that in general we do not expect to see exactly this
distribution at any given time,

but it follows from Theorem~\ref{thm:maxent}
that, after sufficient time,  the distribution is unlikely
to be very far from it.

Does this distribution help us identify the population?
Without further constraints,

it does not.
Say that a distribution of money $M$

(where $M_i$ is the fraction of agents with $i$ dollars)
is \emph{fully supported} if there do not exist

$i$ and $j$ such that $i < j$, $M_j > 0$, and $M_i = 0$.  
As the following lemma shows,
a fully supported distribution can be explained in an infinite
number of different ways.
We take an ``explanation'' of $M$ (which has average amount of money
$m$) to consist of a distribution $\vec{\pi}$
over strategies such
that if a $\pi_k$ fraction of agents use strategy $S_k$ then

$M^*_{\vec{\pi},m} = M$ (i.e., the maximum entropy distribution that
results from those strategies will be $M$).

\lem \label{lem:infinite}
If $M$ is a fully supported distribution of money with finite support,
there there exist an infinite number of explanations of $M$.
\elem

\prf
Fix a value of $\lambda$.  Then $M$ and $\lambda$ determine a
distribution $\vec{\pi}$ as follows.  Suppose that $k$ is the
maximum amount of money that any agent has under $M$ (this is well
defined since the support of $M$ is finite, by assumption).
Then we take
$\pi_k$ to be the unique value that satisfies
$$M_k = \pi_k
\lambda^k / (\sum_{i = 0}^k \lambda^i ).$$

Note that $M_j = \sum_i M_j^i$.  Therefore,

once we have defined $\pi_k$, we can take $\pi_{k-1}$
to be the unique value that

satisfies $$M_{k-1} = \pi_k \lambda^{k-1} / (\sum_{i = 0}^k
\lambda^i ) + \pi_{k-1} \lambda^{k-1} / (\sum_{i = 0}^{k-1} \lambda^i).$$
Iterating this process uniquely defines
$\vec{\pi}$.
However, $\vec{\pi}$ may not be a valid explanation, since
some $\pi_j$ may be negative.
This happens exactly when

$$M_j < \sum_{l = j+1}^k \pi_l \lambda^j / (\sum_{i = 0}^l \lambda^i).$$

As $\lambda$ grows large, the terms on the right side of this

inequality all tend towards 0.  Thus, taking $\lambda$ sufficiently large
ensures that  $\pi_j \ge 0$ for all $j$.
By construction, these values of $\pi_j$ are of a form that satisfied
constraints (\ref{eqn:constraint1}) and (\ref{eqn:constraint2}), so
$\pi$ is a valid explanation for $M$.
Continuing to increase $\lambda$ will give an infinite number of
different explanations.
\eprf

We have not yet

shown that there are types of agents for which the strategies in a given
explanation are the strategies used in equilibrium.
However, by examining the decision problem that
determines the optimal threshold strategy for an agent, it can be
shown that the parameters $\alpha$, $\gamma$, and $\delta$ can be set
so as to make any threshold strategy optimal.

\lem \label{lem:strategies}
Let $M$ be the distribution of money in a nontrivial system and
$\vec{\pi}$ be an explanation for $M$.  Then for
all $\beta > 0$, $\rho > 0$, and $k$, there exist $\alpha$,
$\gamma$, and $\delta$ such that $S_k$ is the best reply
for an agent of type $(\alpha,\beta,\gamma,\delta,\rho)$ to
$\vec{\pi}$.
\elem

\prf (Sketch)
Consider the decision problem faced by an agent when comparing $S_i$
to $S_{i+1}$.
$S_i$ and $S_{i+1}$ differ only in what they do when an agent has $i$
dollars.

As Theorem 4.1 of \cite{scrip06} shows,
to decide whether or not to volunteer if he has $\$i$,
an agent should determine the expected value of having a request
satisfied when he runs out of money if he has $i$ dollars and plays

$S_i$, and volunteer if this value is greater than $\alpha$.
The parameters of the random walk that determines this expectation
are determined by $i$, $M$, and $\vec{\pi}$.
We can make $S_k$ optimal by fixing some $\alpha$ and $\gamma$ and
then adjusting $\delta$ so that not working becomes superior exactly
at $i = k+1$.
\eprf

Lemma \ref{lem:infinite} shows that $M$ has an infinite number of
explanations. Lemma \ref{lem:strategies} shows that we can find an
equilibrium corresponding to each of them.  Given an explanation
$\vec{\pi}$, we can use Lemma \ref{lem:strategies} to find a type
$t_j$ for which strategy $S_j$ with $j$ in the support of $\vec{\pi}$ is
the best
reply to $M$ and $\vec{\pi}$.  Taking $T = \{t_j ~|~ j \in
supp(\vec{\pi}) \}$ and $f_j = \pi_j$ gives us a population for which
$M$ is the equilibrium distribution of money.
This type of population seems unnatural;

it requires one type for each possible amount of money.
We are typically interested in a more parsimonious explanation, one that
requires relatively few types, for reasons the following lemma makes
clear.

\lem \label{lem:uniquemoney}
Let $\vec{\pi}$ be the true explanation for $M$.  If $k$ is the
largest threshold in the support of $\vec{\pi}$ and $s$ is the size of the
support of
$\vec{\pi}$, then any other explanation will have a support of size
at least $k - s$.
\elem

\prf
We know that $M^k_i = \pi_k \lambda^i / \sum_{j = 0}^k \lambda^j$,
where $\lambda$ is the (unique) value that satisfies
constraint (\ref{eqn:constraint1}).  Let $b_k = \pi_k / \sum_{j = 0}^k
\lambda^j$;
then $M_i^k = \lambda^i b_k$, and
$M_i = \sum_k M_i^k = B_i \lambda^i$,
where $B_i = \sum_{k \ge i} b_k$.
Note that if $\pi_{i-1} = 0$,  then $B_i =
B_{i-1}$, so $M_i / M_{i-1} = \lambda$.
Since $s$ strategies get positive probability according to $\vec{\pi}$,
at least $k - s$ of the ratios $M_i/M_{i-1}$ with $1 \le i \le k$
must have value $\lambda$.
Any other explanation will have a different value $\lambda'$
satisfying constraint (\ref{eqn:constraint1}).
This
means that the $k - s$ ratios with value $\lambda$ must correspond to
places where $\pi_i > 0$.  Thus the support of any other explanation
must be at least $k - s$.
\eprf

If $s \ll k$, Lemma \ref{lem:uniquemoney} gives us a strong reason
for preferring the minimal explanation (i.e., the one with the smallest
support); any other explanation will
involve significantly more types.  For $s = 3$ and $k = 50$,
the smallest explanation has a support of at most three thresholds, and
thus requires three types;
the next smallest explanation requires
at least

47 types.  Thus, if the number of types
of agents is relatively small, the minimal explanation will be the
correct one.

\begin{figure}[htb]

\centering \epsfig{file=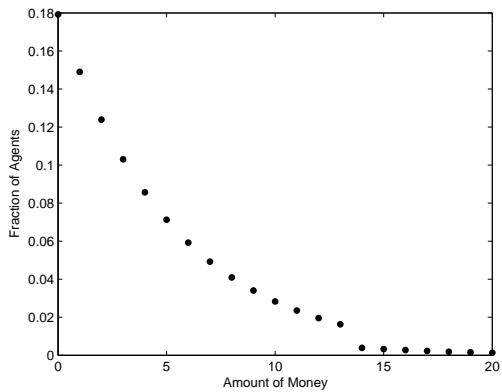, height=2.2in}
\caption{Distribution of money with two types of agents.}
\label{fig:distribution}
\end{figure}

\begin{figure}[htb]

\centering \epsfig{file=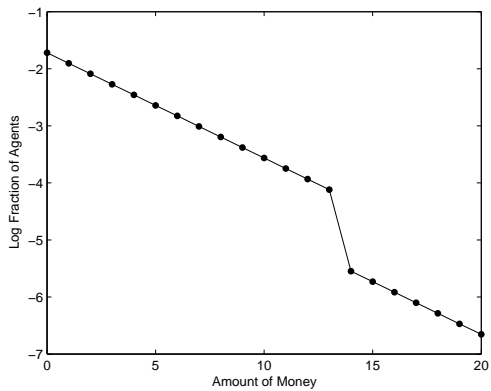, height=2.2in}
\caption{Log of the distribution of money with two types of agents.}
\label{fig:log}
\end{figure}

The proof of Lemma \ref{lem:uniquemoney} also gives us an algorithm
for finding this minimal explanation.  We know that $M_i = B_i
\lambda^i$.  Therefore $\log M_i = \log B_i + i \log \lambda$.  This
means that in a plot of $\log M_i$, ranges of $i$ where $S_i$ is not

played will be a line with slope $\lambda$.  Thus, the minimal
explanation can be found by finding the minimum number of lines of
constant slope that fit the data.  For a simple example of how such a
distribution might look, Figure \ref{fig:distribution} shows an
equilibrium distribution of money for the population
$$(\{(.05,1,1,.95,1),(.15,1,1,.95,1) \},(.3,.7),1000)$$ (i.e., the only
difference between the types is it costs the second type three times
as much to satisfy a request) with $m = 4$ and the equilibrium
strategies $(20,13)$.
Figure \ref{fig:log} has the same distribution plotted on a log
scale.  Note the two lines with the same slope ($\lambda$) and the
break at 14.

Now we have an understanding of how we can take a distribution of
money and infer from it the
minimal explanation of the
number of types of agents, the fraction of
the population composed of each type, and the strategy each type is
playing. (Note that we cannot distinguish multiple types playing the
same strategy.)  We would like to use this information to learn about

the preferences of agents.  The key observation is that the strategy
chosen by an agent will be a best reply to the strategies of the
other agents.
The proof of Lemma \ref{lem:strategies} shows that from $M$ and
$\vec{\pi}$ we can compute the parameters that control the random walk

taken by an agent
playing strategy $S_k$ starting with $k$ dollars until he runs out of
money.  This enables us to compute the expected stopping time of the
random walk and, from this, a type for which $S_k$ is a best reply.
This argument gives us constraints on the set of types for which $S_k$ is
optimal.  These constraints still allow quite a few possible types.
However, suppose that
over time $T$,
the set of types,
remains constant but $\vec{f}$, $n$, and $m$ all vary as
agents join and leave the system.  A later observation with a slightly
different population (but the same $T$) would give another equilibrium
with new constraints on the types of the agents.  A number of
such
observations potentially reveal enough information to

allow strong inferences about agent types.

Thus far we have implicitly assumed that there are only a small number
of types of agents in a system.
Given that a type is defined by five real numbers, it is perhaps more
reasonable to assume that each agent has a different type, but there is
a small number of ``clusters'' of agents with
similar types.
For example, we might believe that generally agents either place a
high value or a low value on receiving service.
While the exact value may vary, the types of two low-value agents
or two high-value agents will be quite similar.
We have also assumed in our analysis that all agents
play their optimal threshold strategy.  However, computing this

optimum may be too difficult for many agents.
Even ignoring computational issues, agents may have
insufficient information about their exact type
or the exact types of other agents to compute the optimal threshold strategy.
Assuming that there are a few clusters of agents with similar, but not
identical, types and/or assuming that agents do not necessarily play
their optimal threshold strategy, but do play a strategy close to
optimal both lead to a similar picture of a system, which is one that we
expect to see in practice:
we will get clusters of agents playing similar strategies (that is,
strategies with thresholds clustered around one value), rather than all

agents in a cluster playing exactly the same strategy.
This change has relatively little impact on our results.

Rather than seeing straight lines representing populations with a
sharp gap between them, as in Figure~\ref{fig:log}, we expect slightly
curved lines representing a cluster of similar populations,

with somewhat smoother transitions.

\section{Optimizing the money supply} \label{sec:optimize}

In Section \ref{sec:distribution} we considered how money is
distributed among agents of different types, assuming that the money
supply is fixed. We now want to examine what happens to the distribution
of money
as the amount of money changes.  In particular, we want to
determine the amount that optimizes the performance of the system. We
will show that increasing the amount of money improves performance
up to a certain point, after which the system experiences a monetary
crash.
Once the system crashes, the only equilibrium will be the trivial one
where all agents play $S_0$.  Thus, optimizing the performance of the
system is a matter of finding out just how much money the system can
handle before it crashes.

Before we can start talking about optimal money supply, we have to
answer a more fundamental question: given an equilibrium, how good is
it?
Consider a single round of the game with a homogeneous population
of some fixed type $t$.  If
a request is satisfied,
social welfare increases by $\gamma_t - \alpha_t$; the requester gains
$\gamma_t$ utility and the satisfier pays a cost of $\alpha_t$.
If no request is satisfied then no utility is gained.  What is the
probability that a request will be satisfied?  This requires two
events to occur.  First, the agent chosen to make a request must have
a dollar, which happens with probability $1 - M_0$.  Second, there
must be a volunteer able and willing to satisfy the request.  Any
agent who does not have his threshold amount of money is willing
to volunteer.  Thus, if $\tau$ is the fraction of agents at their
threshold then the probability of having a volunteer is
$1 - (1 - \beta_t)^{(1 - \tau)n}$.
We believe that in most large systems this probability is quite close
to 1; otherwise, either $\beta_t$ must be unrealistically small or

$\tau$ must be very close to 1.  For example, even if

$\beta = .01$ (i.e., an agent can satisfy 1\% of requests), 1000
agents will be able to satisfy 99.99\% of requests.
If $\tau$ is
close to 1, then agents will have an easier time earning money then
spending money (the probability of spending a dollar is at most $1 /
n$, while for large $\beta$ the probability of earning a dollar if an
agent volunteers is roughly $(1 / n)(1 / (1 - \tau))$).  If an agent
is playing $S_4$ and there are $n$ rounds played a day, this means
that for $\tau = .9$ he would be willing to pay $\alpha$ today to
receive $\gamma$ over 10 years from now.
For most systems, it seems
unreasonable to have $\delta$ or $\gamma / \alpha$ this large.
Thus, for the purposes of our analysis, we

approximate $1 - (1 - \beta_t)^{(1 - \tau)n}$ by 1.

With this approximation, we can
write the expected increase in social welfare each round as
$(1 - M_0)(\gamma_t - \alpha_t)$.  Since we have discount factor $\delta_t$,
the total expected social welfare summed over all rounds is $(1 -
M_0)(\gamma_t - \alpha_t) /
(1 - \delta_t)$.
For heterogeneous types, the situation is essentially the same.  Our

equation for social welfare is more complicated because now the gain in
welfare depends on the $\gamma$, $\alpha$, and $\delta$ of the agents
chosen in that round, but the overall analysis is the same, albeit with
more cases.
Thus our goal is clear: find the amount of money that,
in equilibrium, minimizes $M_0$.

Many of the theorems in Section \ref{sec:theory} begin
``For all $m$ there exist $\delta^*$ and $n^*$ such that if
$(T,\vec{f},n)$ is a payoff-heterogeneous population with $n > n^*$ and
$\delta_t > \delta^*$ for all $t$''.  Intuitively, the theorems require
large $\delta_t$s to ensure that
agents are patient enough that their decisions are dominated by
long-run behavior rather than short-term utility, and large $n$ to
ensure that small changes in the distribution of money do not move it
far from the maximum-entropy distribution.  In the following theorem
and many of our later results, we want to assume 

that
this condition is
satisfied so that we can apply the theorems from
Section~\ref{sec:theory}.  To simplify the
statements of our theorems,
we use ``the standard conditions hold for $m$'' to mean that the
population $(T,\vec{f},n)$ under consideration is such that $n > n^*$
and $\delta_t > \delta^*$ for the $n^*$ and $\delta^*$ that the
theorems require for $m$.

\thm \label{thm:money}

Let $(T,\vec{f},n)$ be a payoff-heterogeneous population such that the
standard conditions hold for $m$ and $m'$, $m < m'$,
and $\vec{k}$ is a nontrivial equilibrium for $(T,\vec{f},n)$ and $m$.
Then if the
average amount of money is $m'$, best-reply dynamics
starting at $\vec{k}$ will converge to some equilibrium $\vec{k}'
\leq \vec{k}$.  Moreover, if $\vec{k}$ is the maximum equilibrium for
$m$, then $\vec{k}'$ is the maximum equilibrium for $m'$.
Furthermore, if $\vec{\pi}(\vec{k})$ is the distribution over
strategies induced by $(T,\vec{f},n)$ and $\vec{k}$,
and
$\vec{k}'$ is a nontrivial equilibrium,
then $(M^*_{\vec{\pi}(\vec{k}'),m'})_0 \leq
(M^*_{\vec{\pi}(\vec{k}),m})_0$.
\ethm

\prf (Sketch)
Suppose that in the
equilibrium with $m$, all agents of type $t$ use
the threshold strategy $S_{k_t}$.   Then
$(M^*_{\vec{\pi}(\vec{k}),m})^k_i =  \lambda_m^i /\sum_{j = 0}^k
\lambda_m^j$, where $\lambda_m$ is the value of that satisfies
constraint
(\ref{eqn:constraint1}) for $m$.
It is relatively straightforward to show
that if $m < m'$, then $\lambda_{m} < \lambda_{m'}$.
If the equilibrium threshold strategy with both $m$ and $m'$ were
the same, then the desired result would be immediate.
Unfortunately, changing
the average amount of money may change the best-reply function.
However, since $\lambda_{m'} > \lambda_m$, it can be shown that
$(M^*_{\vec{\pi}(\vec{k}),m'})^{k_t}_0 \le
(M^*_{\vec{\pi}(\vec{k}),m})^{k_t}_{k_t}$ and
$(M^*_{\vec{\pi}(\vec{k}),m})^{k_t}_{k_t} \ge
(M^*_{\vec{\pi}(\vec{k}),m})^{k_t}_{k_t}$, for all $k$
This increases the probability of an agent earning a dollar,
so the best reply can only decrease.

Thus, $\BR_{(T,\vec{f}),m'}(\vec{k}) \leq
\vec{k}$.
Applying best-reply dynamics to $\BR_{(T,\vec{f}),m'}$ starting at $\vec{k}$,
as in Theorem~\ref{thm:equilib},
gives us an equilibrium
$\vec{k}'$ such that $\vec{k}' \le \vec{k}$.
Decreasing strategies only

serves to further decrease 
$(M^*_{\vec{\pi}(\vec{k}'),m'})^{k_t'}_0$, so as long as $\vec{k}$ is
nontrivial we will have $(M^*_{\vec{\pi}(\vec{k}'),m'})_0 \leq
(M^*_{\vec{\pi}(\vec{k}),'})_0$.
\eprf

Theorem \ref{thm:money} makes several strong statements about what
happens to social welfare as the amount of money increases (assuming
there is no monetary crash).  Taking the worst-case view, we know
social welfare at the maximum equilibrium will increase.
Alternatively, we can think of the system as being jolted out of
equilibrium by a sudden addition of money.  If agents react to this
using best-reply dynamics and find a new nontrivial equilibrium,
social welfare will have increased.
In general, Theorem~\ref{thm:money}
suggests that, as long as
nontrivial equilibria exist, the more money the better.
As the following theorem shows, increasing the amount of money

sufficiently leads to a monetary crash; moreover, once the system
has crashed, adding more money does not make things better.

\cor \label{cor:crash}
If  $(T,\vec{f},n)$ is a payoff-heterogeneous population  for which the
standard conditions hold for $m$, then
there exists a finite threshold
$m_{T,\vec{f}}$ such that there exists a nontrivial equilibrium if the
average amount of money is
less than $m_{T,\vec{f}}$ and there does not exist a nontrivial
equilibrium
if the average amount of money is greater than $m_{T,\vec{f}}$.
(A nontrivial equilibrium may or may not exist if the average amount of

money is exactly $m_{T,\vec{f}}$.)
\ecor

\prf
Fix $(T,\vec{f},n)$.  To see that there is some average amount of money
$m$ for which there is no nontrivial equilibrium in this population,
consider any average amount $m$.  If there is no nontrivial equilibrium
with $m$, then we are done.  Otherwise, suppose the maximum equilibrium

with $m$ is $\vec{k}_m$.  Let $\vec{k}$ be such that
$\BR^t_{T,\vec{f},m}(\infty, \ldots, \infty) = k_t$.  We must have
$\vec{k}_m \le \vec{k}$.

Choose  $m' >  \sum_t f_t k_t$.

Suppose that the maximum equilibrium with $m'$ is
$\vec{k}_{m'}$.  By Theorem~\ref{thm:money}, we must have
$\vec{k}_{m'} \le
\vec{k}_m$.  Thus, $\vec{k}_{m'} \le \vec{k}$. But if $\vec{k}_{m'}$ is
a nontrivial equilibrium, then in equilibrium, each agent of type $t$
has at most $k_t$ dollars, so the average amount of money in the system
is at most

$\sum_t f_t k_t < m'$.
Thus, there cannot be a nontrivial
equilibrium if the average amount of money is $m'$.

Let $m_{T,\vec{f}}$ be the infimum over all $m$ for
which no nontrivial equilibrium exists with population $(T,\vec{f},n)$
if the average amount of money is $m$.
Clearly, by choice of $m_{T,\vec{f}}$, if $m < m_{T,\vec{f}}$, there is
a nontrivial equilibrium.  Now suppose that
$m > m_{T,\vec{f}}$.  By the construction of
$m_{T,\vec{f}}$, there exists $m'$ with $m > m' > m_{T,\vec{f}}$ such
that no
nontrivial equilibrium exists if the average amount of money is $m'$.
Suppose, by way of contradiction, that there exists a nontrivial
equilibrium if the average amount of money is $m$.  Suppose that the
maximum equilibrium is $\vec{k}$.  By the same argument as that used
in Theorem~\ref{thm:money}, the maximum equilibrium $\vec{k}'$ if the
average
amount of money is $m'$ is such that $\vec{k}' > \vec{k}$.  Thus,
$\vec{k}'$ is a nontrivial equilibrium.
\eprf

\begin{figure}[htb]

\centering \epsfig{file=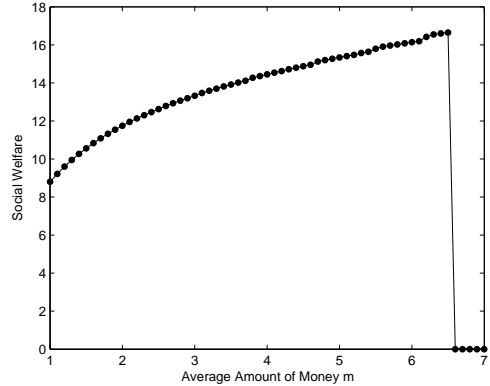, height=2.2in}
\caption{Social welfare for various average amounts of money,
demonstrating a monetary crash.}
\label{fig:crash}
\end{figure}

Figure \ref{fig:crash} shows an example of the monetary crash in a
system with the same population used in Figure

\ref{fig:distribution}.  Equilibria were found using best-reply dynamics
starting at $(100,100)$.

In light of Corollary~\ref{cor:crash}, the system
designer should try to find $m_{T,\vec{f}}$.  How can he do this?
If he knows $T$ and $\vec{f}$, then he can perform a binary search for
$m_{T,\vec{f}}$ by choosing values of $m$ and then using the algorithm from
Section \ref{sec:theory} to see if a nontrivial equilibrium exists.
Observing the system over time and using the techniques described in
Section \ref{sec:distribution} along with additional information he
has about the system may be enough to make this a practical option.

We expect a monetary crash to be a real phenomenon in a system where
the price of servicing a request is fixed.  This can be the case in

practice, as shown by in the babysitting co-op example.  If
the price can change, we expect that as the money supply increases,

there will be inflation; the price will increase so as to avoid a
crash.  However, floating prices can create other monetary problems,
such as speculations, booms, and busts.

Floating prices also impose transaction costs on agents.  In systems
where prices would normally be relatively stable, these transaction
costs may well outweigh the benefits of floating prices.

\section{Altruists and Hoarders} \label{sec:altruists}

Thus far, we have considered standard agents with a type
of the form $(\alpha, \beta, \gamma, \delta, \rho)$.
We have a fairly complete picture of what happens in a system with
only standard agents: increasing the money supply increases agent
utility until a certain threshold is passed and the system has a
monetary crash.
However, any real system will have agents that, from perspective of
the designer, behave oddly.

These agents may be behaving irrationally or they may simply have a
different utility function. 
For our results to be useful in practice, we need to understand how they
are affected by such nonstandard agents.  We consider here two such
types of nonstandard agents, both of which have been observed in real
systems: altruists and hoarders.
Altruists, who provide service without requiring payment, reduce the
incentive for standard agents to work.  The end result is a decrease
in the equilibrium threshold for standard agents.
As a result, an excess of altruists, like an excess of money, can cause
standard agents to stop being willing to work.
However, up to the point where the system has a monetary crash,
altruists improve the
utility of
standard agents.

Hoarders, who want to collect as much money as
possible whether it is actually useful or not,
can be modeled as playing
$S_\infty$.  Since hoarders effectively absorb money, they make the
remaining money more valuable, which increases
the threshold used by standard agents in equilibrium.
This results in reduced utility for standard agents, provided that the
amount of money in the system is constant.

Altruists may at first appear purely beneficial to standard agents,
since they satisfy requests with no cost to standard agents.
Indeed, as the following theorem shows, as long as the system does not
have a monetary crash, altruists do make life better for standard
agents.
To show this, we assume that a fraction $a$ of requests get
satisfied at no cost.  Intuitively, these are the
requests satisfied by the altruists, although
the following result also applies to
any setting where agents occasionally have a (free) outside option.

\thm \label{thm:altruists}

Consider a homogeneous population and assume that the standard conditions
hold for $m$.  Suppose that a fraction $a$ of requests can be satisfied
at no cost.  Then as long as the system does not have a
monetary crash, social welfare increases as $a$ increases (assuming

that
the maximum equilibrium is always played).
\ethm

\prf (Sketch)
As we discussed in Section \ref{sec:optimize}, the expected social
welfare with a single type of standard agent is $(1 - M_0)(\gamma -
\alpha) / (1 - \delta)$.
Rounds where a request is satisfied at no cost generate
social welfare of $a \gamma / (1 - \delta)$.  The
same analysis as in Section \ref{sec:optimize} shows that the
remaining rounds generate social welfare of
$(1-a)(1 - M_0(a))(\gamma - \alpha) / (1 - \delta)$,
where $M_0(a)$ is the new equilibrium value of $M_0$ given that
a fraction $a$ of requests are satisfied at no cost.
Thus, the total welfare as a function of $a$ is
$$a \gamma / (1 - \delta) + (1-a)(1 - M_0(a))(\gamma - \alpha) / (1 -
\delta).$$
To see that this increases as $a$ increases, we would like to
take the derivative relative to $a$ to get $(\gamma
+ (\gamma - \alpha)(- M_0'(a)(1 - a) - (1 - M_0(a)))) / (1 -
\delta)$.
Unfortunately $M_0(a)$ may not even be continuous.  Because strategies
are integers, there will be regions where $M_0(a)$ is constant, and
then a jump when a critical value of $a$ is reached that causes the
equilibrium to change.  In those regions where $M_0(a)$ is constant,
$M_0'(a)$ is 0.  Since
$1 / (1 - \delta)(\gamma
+ (\gamma - \alpha)(- (1 - M_0(a))))$ is positive, social welfare
is increasing in these regions.  At points where $M_0(a)$ jumps, the
equilibrium strategies will decrease.   The proof of
Theorem \ref{thm:money} shows that this decreases the value of $M_0(a)$
(unless
the system crashes).  This
means that social welfare also increases at these points, so social
welfare will always increase unless the system crashes.
\eprf

Theorem~\ref{thm:altruists} shows that altruists do not hurt provided
that there are not enough them to crash the monetary system.  But what
if there are enough to crash the monetary system?  As Adar and
Huberman \citeyear{adar00} observed, this is essentially
what has happened in Gnutella, where 70\% of agents do not share any
files, and nearly 50\%  of responses are from the top 1\%
of sharing hosts.
In cases like this, whether standard agents are better off depends on

the parameters of the system; the altruists may be satisfying fewer
requests than would be satisfied in a cooperative equilibrium, but
they are also not imposing the cost of satisfying these requests on
the standard agents.

Turning to hoarders, it seems in practice that
whenever a system allows agents to accumulate something, be it work
done, as in Seti@home, friends on online social networking sites, or
``gold" in an online game, a certain group of agents seems to make it
their goal to accumulate as much of it as possible.  In pursuit of
this, they will engage in behavior that seems
irrational.
For simplicity here, we model hoarders as playing the strategy
$S_\infty$.  This means that they will volunteer under all
circumstances.  Our analysis would not change significantly if we also
required that they never made a request for work.  Our first result
shows that, for a fixed money supply, having more hoarders makes
standard agents worse off.

\thm\label{thm:hoarders}
Suppose that the standard agents in a system are described by

$(T,\vec{f})$.  Let
$f_H$ be the fraction of hoarders (i.e., with probability
$f_H$ an agent is a hoarder and with probability $1 - f_H$ he is

described by $(T,\vec{f},n)$).  Assume that $(T,\vec{f},n)$
satisfies the standard conditions

for $m$.  Then, as $f_H$ increases, the utility of standard agents
in the maximum equilibrium decreases.
\ethm

\prf (Sketch)
Increasing $f_H$ is equivalent to taking some number of standard
agents and increasing their strategy to $S_\infty$.  Lemma
\ref{lem:monotone} shows us that the thresholds in equilibrium do not
decrease.  It can be shown by examining the formula for the
maximum-entropy distribution that increasing strategies increases $M_0$.
As
the discussion in Section \ref{sec:optimize} shows, this will decrease
agent utility.
\eprf

Both Theorem~\ref{thm:altruists} and~\ref{thm:hoarders} assume that the
average amount of money in the system
remains constant.
However, the actions of both hoarders and altruists are visible to a
system designer.  The designer can take their existence into
account when setting the money supply.  As a general guideline,
altruists require that money be removed from the system while hoarders
require that it be added.

Hoarders also have a beneficial aspect.  As we have observed, a monetary
crash occurs when a dollar becomes valueless because there
are no agents willing to take it.  However, with hoarders in the system,
there is always someone to take it.
Suppose that every standard agent in a system with hoarders plays $S_0$.
As we saw in Lemma \ref{lem:strategies}, we can compute the
expected utility for having a single dollar and playing $S_1$.  As
long as this utility is greater than $\alpha$, the best reply will be
greater than $S_0$.  From Theorem \ref{thm:equilib},  it follows that
there will be a nontrivial equilibrium.  Therefore, by Theorem
\ref{thm:money},  increasing $m$ will always increase social
welfare. In practice, this seems like an unrealistic outcome.
It seems likely that hoarders hoard only if money is seen

as a scarce resource. Thus, as a practical matter, there is likely to
be a limit to how far the system designer can increase $m$.

\section{Conclusions and Future Work} \label{sec:conclusion}

For our model of a scrip system with payoff-heterogeneous types, we
have proved that equilibria exist in threshold
strategies and that the distribution of money in these equilibria is
given by maximum entropy.  As part of our equilibrium argument, we
showed that the best-reply function is monotone.  This proves the
existence of equilibria in pure strategies and permits efficient
algorithms to compute these equilibria.  We have also examined some
of the practical implications of these theoretical results.  For
someone interested in studying the agents of scrip systems, our
characterization of equilibrium distribution of money forms the basis
for techniques relevant to inferring characteristics of the agents of a
scrip system from the distribution of money.   For a system designer,
our results on optimizing the money supply provide a simple maxim:
keep adding money until the system is about to experience a monetary
crash.  Finally, we
provide insight into the effects of altruists and hoarders on a scrip

system, as well as providing guidance to system designers for dealing
with them.

There remain a number of interesting areas for future work,
including the following:
\begin{itemize}
\item
Our theoretical results all assume that agents are merely payoff

heterogeneous.  Varying $\beta$ or the request rate poses a problem for
our techniques because the stationary distribution of the Markov chain

will no longer be uniform, so we can no longer describe the equilibrium
distribution of money using maximum entropy.  However, there is
nothing about this type of heterogeneity that intuitively poses a
problem for convergence or threshold strategies.  Indeed, we would
expect that, knowing the average amount of money for one particular

type, the distribution of money within that type should still follow a
maximum-entropy distribution, because of the uniformity within the
type.  This would lead to a distribution of money that has a similar

form, although now each type would have a different value of
$\lambda$.  If this is the case, many of our results still hold.
\item
It seems unlikely that altruism and hoarding are the only two
types of ``irrational'' behavior we will find in real systems.  Are there
other major types that our model can provide insight into?
\item
It seems natural that the behavior of a very small group of agents
should not be able to change the overall behavior of the system.  Can
we prove results about equilibria and utility when a small group
follows an arbitrary strategy?  This is particularly relevant when

modeling attackers.  See \cite{ADGH06} for general results in this setting.
\item
Our model makes a number of strong predictions about the agent
strategies, distribution of money, and effects of variations in the
money supply.  It also provides techniques to help analyze
characteristics of agents of a scrip system.  It would be interesting
to test these predictions on a real functioning scrip system to either
validate the model or gain insight from where its predictions are
incorrect.

\item What happens if we allow prices to vary over time, and have
$\alpha$ and $\gamma$ varying by round

as in the work of Hens et al. \cite {hens}.
In this setting threshold strategies still seem natural to determine
whether to work, but now the threshold would have to  be a function of
$\alpha$ and the payment that would be received rather than a
constant.  Furthermore, agents may wish to not make a request at all
(even if they have the money to pay for it) if $\gamma$ is too small or
the cost is too large.  Again, we would expect to see thresholding
behavior in these strategies.
\end{itemize}

\subsection*{Acknowledgements}

We would like to thank Dan Reeves, Randy Farmer, and others at Yahoo!
research for helpful discussions about applications of scrip systems

and three anonymous referees for helpful suggestions.

 EF, IK
and JH are supported in part by NSF under grant ITR-0325453.  JH is

also supported in part by NSF under grant
IIS-0534064

and by AFOSR under grant

FA9550-05-1-0055.

\end{document}